# ANALYTICAL TREATMENT OF PARTICLE-CORE INTERACTION

Yuri K. Batygin

Los Alamos National Laboratory, Los Alamos, NM 87544

## Abstract

Particle-core interaction is the well-developed model of halo formation in high-intensity beams. In present paper an analytical solution for averaged single particle dynamics around uniformly charged beam core is obtained. The problem is analyzed through sequence of canonical transformations of Hamiltonian describing nonlinear particle oscillations. An analytical expression for maximum particle deviation from the axis is obtained. Results of the study are in good agreement with numerical simulations and with previously achieved data.

## 1. Introduction

Injection of non-uniform beam into linear focusing channel results in beam emittance growth and appearance of small fraction of particles outside of the beam core. Halo particles perform large-amplitude oscillations around beam core and can be lost at the aperture of the channel. Activation of the accelerating channel by randomly lost particles is a serious problem of operation of the high-intensity accelerators. Mechanism of halo formation was intensively studied through particle-core model [1-5].

## 2. Particle-core model

The model is based on simultaneous solution of equation for oscillation of the beam envelope $r = R/R_o$ around equilibrium beam radius $R_o$, and equation for particle deviation from the axis $u=x/R_o$:

$$\frac{d^2r}{d\tau^2} + r - \frac{1}{(1+b)r^3} - \frac{b}{(1+b)r} = 0, \qquad (1)$$

$$\frac{d^2u}{d\tau^2} + u = \begin{cases} \dfrac{b}{(1+b)}\dfrac{u}{r^2}, & |u| \leq r \\ \dfrac{b}{(1+b)u}, & |u| > r \end{cases}, \qquad (2)$$

where $\tau = \Omega_r t$ is the dimensionless time, $\Omega_r$ is the frequency of particle oscillation in uniform focusing channel, $b$ is the dimensionless space charge parameter

$$b = \frac{2}{\beta\gamma}\frac{I}{I_c}\frac{R_o^2}{\varepsilon^2}, \qquad (3)$$

$I$ is the beam current, $I_c = 4\pi\varepsilon_o mc^3/q$ is the characteristic beam current, $\varepsilon$ is the normalized beam emittance, $\beta$ is the particles velocity, $\gamma$ is the particle energy. In particle-core model, beam affects single particle motion through beam envelope appearing in Eq. (2), while particle does not affect beam envelope. Fraction of particles populating the halo is small, and their effect on beam core can be neglected.

Fig. 1 illustrates typical single-particle trajectories around oscillating core resulting from numerical integration of Eqs. (1), (2). Depending on initial conditions, particle can oscillate (a) inside the core, (b) around the core, (c) far from core with constant amplitude, and (d) resonantly from the core. Fig. 2 illustrates the same trajectories as stroboscopic images in phase space, where particle positions are taken after each second core oscillation period. In present paper we will find analytical solution for averaged particle trajectories in particle-core model.

Beam envelope performs oscillations in continuous focusing channel around equilibrium value of $r = 1$ (see Appendix)

$$r = 1 + \Delta \cos(2\Omega\tau), \qquad (4)$$

where envelope oscillation frequency is

$$2\Omega = \sqrt{2(\frac{2+b}{1+b})}. \qquad (5)$$

Space charge field of the uniform beam consists of linear oscillating part and of non-oscillating nonlinear part, see Eq. (2). Following Ref. [1], let us approximate nonlinear part by cubic term (see Fig. 3). With that approximation, equation of particle motion becomes:

$$\frac{d^2u}{d\tau^2} + u - (\frac{b}{1+b})\{\frac{u}{[1+\Delta\cos(2\Omega\tau)]^2} - \frac{u^3}{4}\} = 0. \qquad (6)$$

Then, using expansion

$$\frac{1}{(1+\Delta\cos 2\Omega\tau)^2} \approx 1 - 2\Delta\cos 2\Omega\tau, \qquad (7)$$

equation of particle motion is:

$$\frac{d^2u}{d\tau^2} + u(\frac{1}{1+b})(1 + 2\Delta b \cos 2\Omega\tau) + (\frac{b}{1+b})\frac{u^3}{4} = 0. \tag{8}$$

Equation (8) corresponds to Hamiltonian:

$$H = \frac{\dot{u}^2}{2} + \varpi^2 \frac{u^2}{2}(1 - h\cos 2\Omega\tau) + \alpha\frac{u^4}{4}, \tag{9}$$

where the following notations are used:

$$\varpi^2 = \frac{1}{1+b}, \qquad h = -2b\Delta, \qquad \alpha = \frac{b}{4(1+b)}. \tag{10}$$

Hamiltonian, Eq. (9), describes inharmonic oscillator with parametric excitation. Presence of nonlinear term $\sim u^4$ limits amplitude of particle oscillation around the core.

### 3. Averaged Hamiltonian

Let us change variables $(\dot{u}, u)$ in Hamiltonian, Eq. (9), for new variables $(Q, P)$ using generating function:

$$F_2(u,P,\tau) = \frac{uP}{\cos\Omega\tau} - (\frac{P^2}{2\varpi} + \varpi\frac{u^2}{2})tg\Omega\tau. \tag{11}$$

Relationships between variables are given by:

$$\begin{cases} Q = \frac{\partial F_2}{\partial P} = \frac{u}{\cos\Omega\tau} - \frac{P}{\varpi}tg\Omega\tau \\ \dot{u} = \frac{\partial F_2}{\partial u} = \frac{P}{\cos\Omega\tau} - u\varpi\, tg\Omega\tau \end{cases}, \tag{12}$$

which results in the following transformation from old variables to new variables [6]:

$$\begin{cases} u = Q \cos \Omega \tau + \dfrac{P}{\varpi} \sin \Omega \tau \\ \dot{u} = -\varpi Q \sin \Omega \tau + P \cos \Omega \tau \end{cases}. \tag{13}$$

Transformation (12) is a representation of solution of equation of a single particle motion as a combination of fast oscillating terms with frequency $\Omega$ and slow variables $Q, P$. New Hamiltonian is given by $K = H + \dfrac{\partial F_2}{\partial \tau}$:

$$K = \frac{P^2}{2} + \varpi^2 \frac{Q^2}{2} - \frac{\varpi^2 h}{2} (Q \cos \Omega \tau + \frac{P}{\varpi} \sin \Omega \tau)^2 \cos 2\Omega \tau$$

$$+ \frac{\alpha}{4} (Q \cos \Omega \tau + \frac{P}{\varpi} \sin \Omega \tau)^4 - \frac{P^2 \Omega}{2\varpi} - \frac{\Omega \varpi}{2} Q^2. \tag{14}$$

After averaging all time-dependent terms over period of $T = 2\pi/\Omega$ ($<\cos^4 \Omega \tau> = \dfrac{3}{8}$, $<\cos^2 \Omega \tau \sin^2 \Omega \tau> = \dfrac{1}{8}$, $<\cos^3 \Omega \tau \sin \Omega \tau> = 0$, etc.) the Hamiltonian, Eq. (14), becomes:

$$\bar{K} = \frac{\varpi^2 \bar{Q}^2}{2} (1 - \frac{\Omega}{\varpi} - \frac{h}{4}) + \frac{\bar{P}^2}{2} (1 - \frac{\Omega}{\varpi} + \frac{h}{4}) + \frac{3}{32} \alpha (\bar{Q}^2 + \frac{\bar{P}^2}{\varpi^2})^2, \tag{15}$$

where $\bar{Q}, \bar{P}$ are average values of variables at the period $T$. Let us make an additional canonical transformation and change variables $(\bar{Q}, \bar{P})$ for action-angle variables $(J, \psi)$ utilizing generating function

$$F_1(\bar{Q}, \psi) = \frac{\varpi \bar{Q}^2}{2 \text{tg} \psi}. \tag{16}$$

Transformation is given by:

$$\begin{cases} \bar{Q} = \sqrt{\dfrac{2J}{\varpi}} \sin \psi \\ \bar{P} = \sqrt{2J\varpi} \cos \psi \end{cases}. \tag{17}$$

After substitution of transformation, Eq. (17), into Eq. (15), the new Hamiltonian is:

$$\bar{K} = \upsilon J + \kappa J^2 + 2\chi J \cos 2\psi, \qquad (18)$$

with the following notations:

$$\upsilon = \varpi - \Omega = \frac{\sqrt{2} - \sqrt{2+b}}{\sqrt{2(1+b)}}, \quad \kappa = \frac{3}{8}\frac{\alpha}{\varpi^2} = \frac{3}{32}b, \quad \chi = \varpi\frac{h}{8} = -\frac{1}{4}\frac{b\Delta}{\sqrt{1+b}}. \qquad (19)$$

Hamiltonian, Eq. (18), describes excitation of nonlinear parametric resonance. Phase space trajectories, corresponding to constant value of Hamiltonian (18) are presented in Fig. 4. There are two types of separatrices at phase plane. Inner separatrix corresponds to particle motion around the beam core. Another two separatrices correspond to resonant particle oscillations outside of beam core. Presence of nonlinear term limits maximum amplitude of particle oscillations. Width of the resonance defines maximum deviation of particle from the axis.

Fig. 5 illustrates comparison of obtained analytical solution with results of numerical integration of particle trajectories. For low intensity beam, $b = 0.1$, analytical solution is close to numerical integration of particle trajectory. In low intensity regime, ratio of envelope oscillation frequency to that of particle is close to factor of 2, and conditions of canonical transformation, Eq. (13), are fulfilled properly. With increase of beam intensity, this ratio is out of rational value, which makes suggestion of selected solution, Eq. (13), less applicable. However, even for high intensity beam, the maximum deviation of particle from the axis described by analytical solution is close to that obtained by numerical integration.

## 4. Nonlinear Parametric Resonance

Let us determine the maximum deviation of particle from the axis. There are certain specific unmovable points (stable, $J_s$, and unstable, $J_u$) at phase plane of averaged motion, see Fig. 4. Equations of motion for unmovable points follow from Hamiltonian, Eq. (18):

$$\frac{dJ}{d\tau} = -\frac{\partial \bar{K}}{\partial \psi} = 4\chi J \sin 2\psi = 0 , \quad (20)$$

$$\frac{d\psi}{d\tau} = \frac{\partial \bar{K}}{\partial J} = \upsilon + 2\kappa J + 2\chi \cos 2\psi = 0 . \quad (21)$$

First equation has a solution $sin 2\psi = 0$. From this condition, unstable points are determined by equations

$$\cos 2\psi = 1, \qquad \psi = 0, \pi, \qquad J_u = -\frac{\upsilon + 2\chi}{2\kappa}, \quad (22)$$

while stable points are determined as

$$\cos 2\psi = -1, \qquad \psi = \frac{\pi}{2}, \frac{3\pi}{2}, \qquad J_s = \frac{-\upsilon + 2\chi}{2\kappa} . \quad (23)$$

Particle with initial conditions, close to unstable point $J_u$, can perform either small amplitude oscillations inside the inner separatrices, or large - amplitude oscillations with maximum amplitude up to $J_{max}$. Let us define the value of $J_{max}$.

The value of Hamiltonian is the same at the internal and at the outer separatrices. At the internal separatrix, the value of Hamiltonian $\bar{K}_u = \bar{K}(J_u)$ from Eqs. (18), (22) is

$$\bar{K}_u = -\frac{(\upsilon + 2\chi)^2}{4\kappa} . \quad (24)$$

The outer separatrix touches the inner one at the unstable point $J_u$. Particle with the value of Hamiltonian $\bar{K}(J_u)$ can reach the point $J_{max}$ having $\psi=\pi/2$. The value of $J_{max}$ is defined by substitution $\bar{K} = \bar{K}_u$ and $\cos 2\psi = -1$ into Hamiltonian equation (18):

$$\kappa J_{max}^2 + J_{max}(\upsilon - 2\chi) - \bar{K}_u = 0. \tag{25}$$

Eq. (25) has a solution for $J_{max}$:

$$J_{max} = \frac{(-\upsilon + 2\chi) + \sqrt{8|\upsilon\chi|}}{2\kappa}. \tag{26}$$

The value of $J$ is connected with variables $(u, du/d\tau)$ via Eqs. (13), (17) as:

$$J = \frac{1}{2}(u^2\varpi + \frac{\dot{u}^2}{\varpi}). \tag{27}$$

Maximum value of particle position from Eq. (27) is therefore $u_{max} = \sqrt{\dfrac{2J_{max}}{\varpi}}$, or:

$$\frac{x_{max}}{R_o} = \sqrt{\frac{32}{3}\frac{\sqrt{1+\dfrac{b}{2}}-1+\dfrac{b|\Delta|}{2}+\sqrt{2b|\Delta|(\sqrt{1+\dfrac{b}{2}}-1)}}{b}}. \tag{28}$$

In Figs. 6,7 the value of $u_{max}$ as a function of envelope amplitude $\Delta$ for different values of beam space charge parameter $b$ are presented. Fig. 6 illustrates comparison of analytical results, Eq. (28), and results of numerical solution of Eqs. (1), (2). This dependencies show that maximum deviation of particle from the axis can be several times larger than the beam core size. For comparison, Fig. 7 contains also empirical dependence

$$\frac{x_{max}}{(R_o/2)} = A + B\ln(\mu), \tag{29}$$

where $A = B = 4$, $\mu = 1+\Delta$, obtained in Ref. [3] as a generalization of numerical and experimental data. Comparison shows good agreement between analytical predictions, Eq. (28), and previously obtained data.

## 5. Conclusion

Particle-core model was treated analytically through Hamiltonian canonical transformations. Resulted equations describe averaged single particle trajectories around the core. Analytical expression for maximum particle deviation from the axis is obtained. Results are in good agreement with numerical simulations and with previously achieved data.

**Appendix**

Equation for dimensionless beam envelope $r = R/R_o$

$$\frac{d^2r}{d\tau^2} + r - \frac{1}{(1+b)r^3} - \frac{b}{(1+b)r} = 0, \quad (A1)$$

has the equilibrium solution $r = 1$. Consider small deviation of beam radius from equilibrium condition

$$r = 1 + \vartheta. \quad (A2)$$

Substitution of expansions

$$\frac{1}{r} \approx 1 - \vartheta, \qquad \frac{1}{r^3} \approx 1 - 3\vartheta, \quad (A3)$$

into envelope equation (A1) gives for small deviation from the equilibrium:

$$\frac{d^2\vartheta}{d\tau^2} + 2(\frac{2+b}{1+b})\vartheta = 0. \quad (A4)$$

Equation (A4) has the solution

$$\vartheta = \Delta \cos(\sqrt{2(\frac{2+b}{1+b})}\tau + \psi_0). \quad (A5)$$

For small intense beam $b \approx 0$, envelope oscillates with the frequency close to double of that of a single particle within the beam core:

$$\vartheta = \Delta \cos(2\tau + \psi_o), \quad (A6)$$

while for high intensity beam, $b \gg 1$, envelope oscillation frequency is a factor of $\sqrt{2}$ smaller:

$$\vartheta = \Delta \cos(\sqrt{2}\tau + \psi_o). \quad (A7)$$

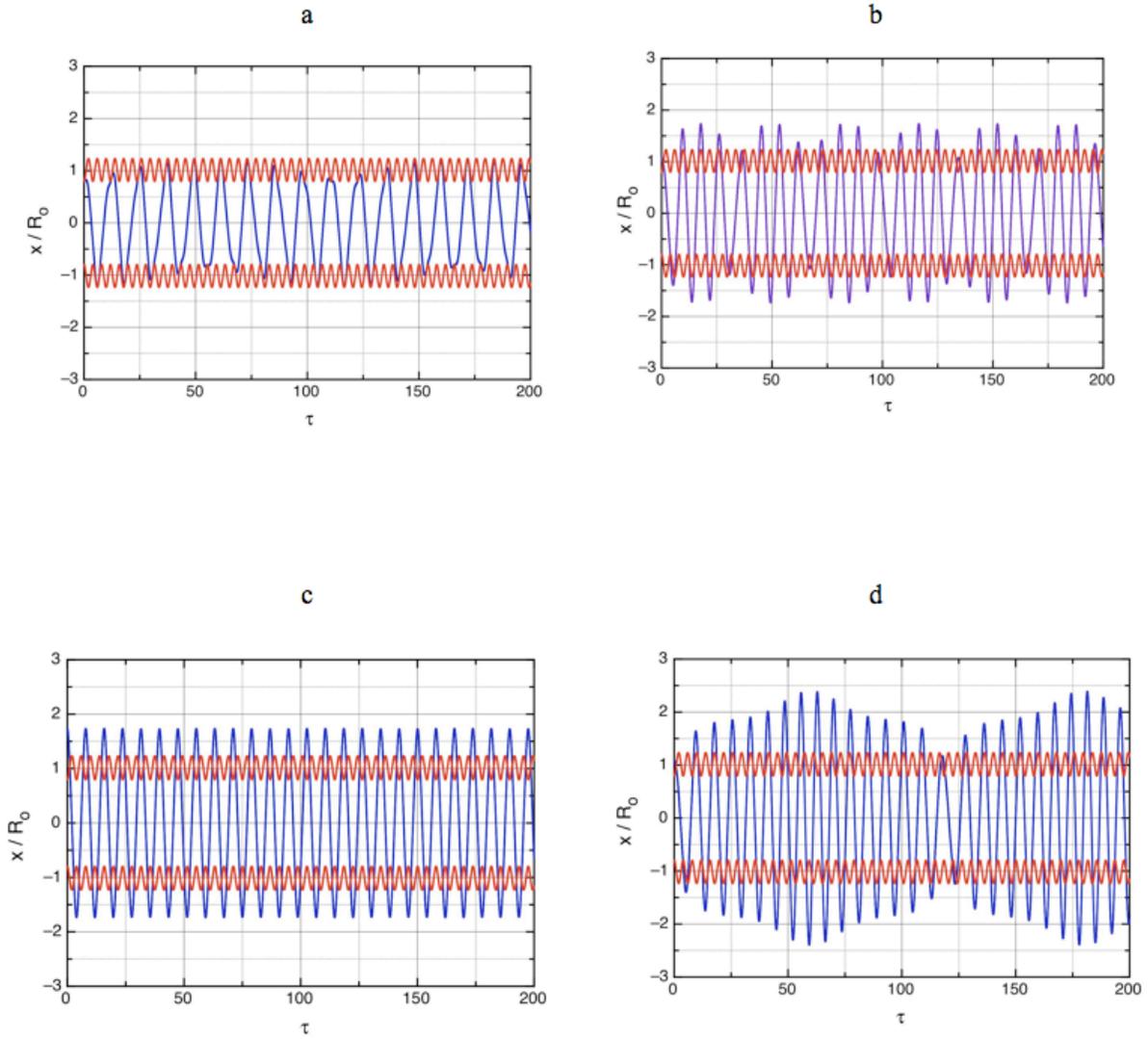

Fig. 1. Envelope oscillations of the beam with space charge parameter $b = 3$, amplitude $\Delta = 0.2$, and single particle trajectories with initial conditions (a) $x_o/R_o=0.8$, (b) $x_o/R_o =1.071$, (c) $x_o/R_o =1.728$, (d) $x_o/R_o =1.082$.

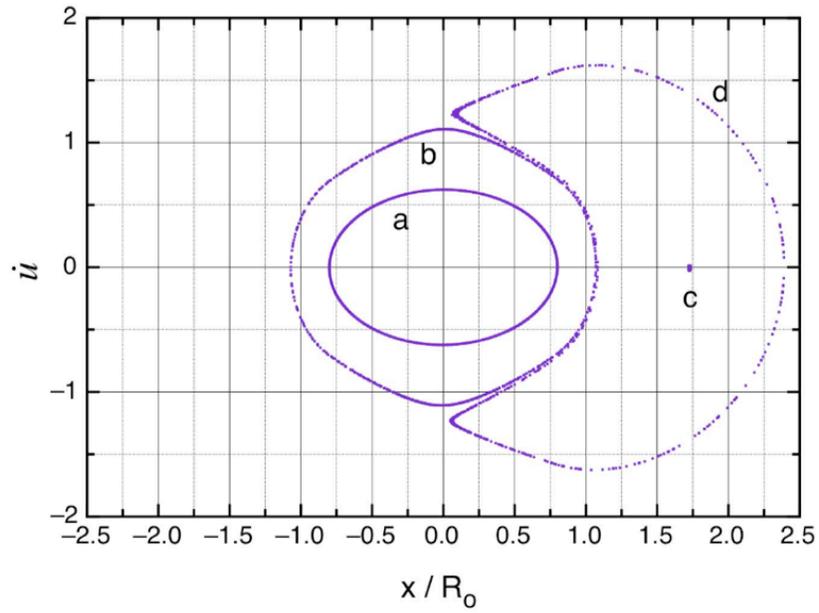

Fig. 2. Stroboscopic particle trajectories at phase plane ($u$, $du/d\tau$) taken after each two envelope oscillation periods: (a) $x_o/R_o$=0.8, (b) $x_o/R_o$ =1.071, (c) $x_o/R_o$ =1.728, (d) $x_o/R_o$ =1.082.

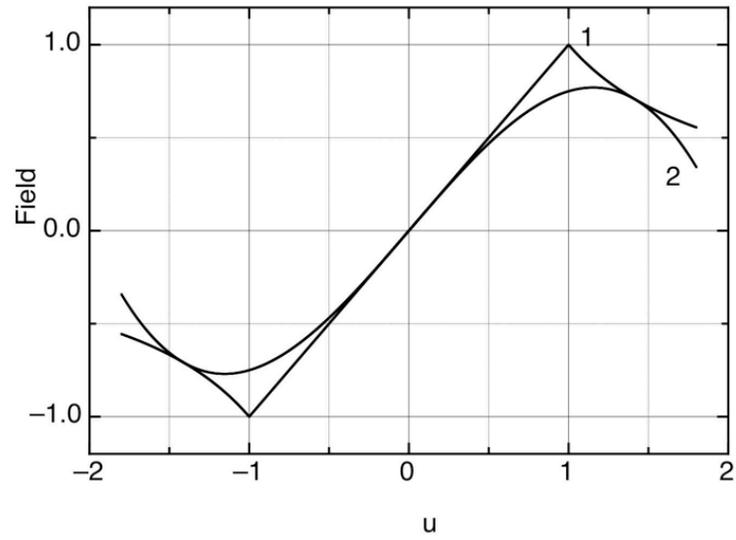

Fig. 3. (1) Space charge field of the uniformly charged core, and (2) field approximation in Eq. (6).

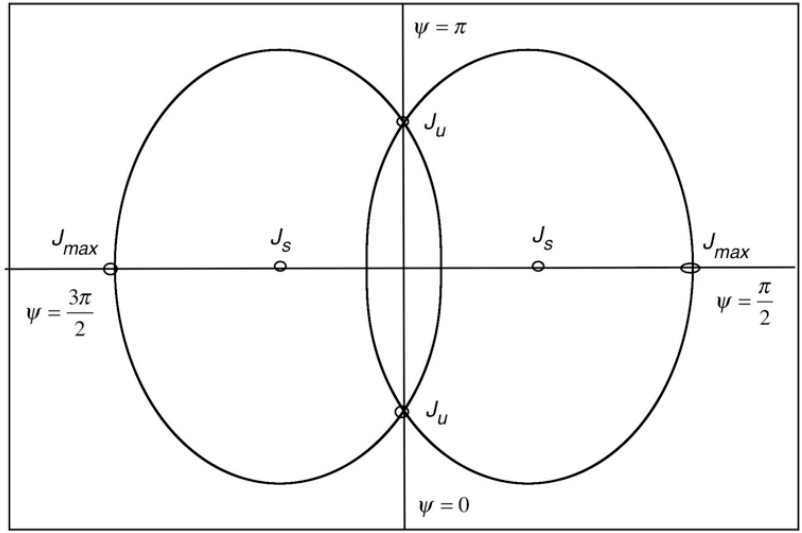

Fig. 4. Phase space trajectories of averaged particle motion around beam core, Eq. (18).

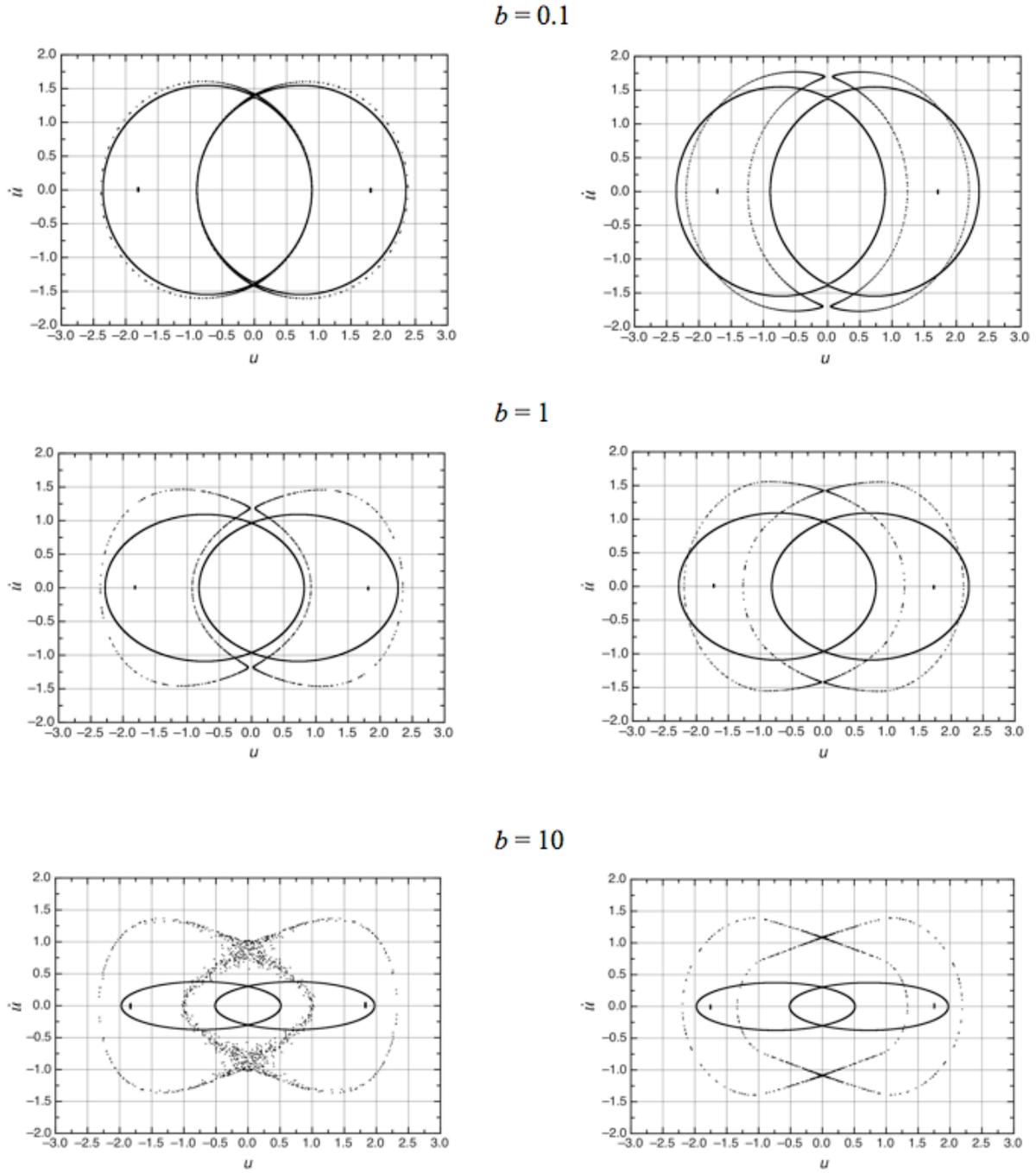

Fig. 5. (Solid) analytical and (dotted) numerical results of averaged phase space trajectories in (left) approximate field, Eq. (6), and (right) in space charge field of the uniform core, Eq. (2).

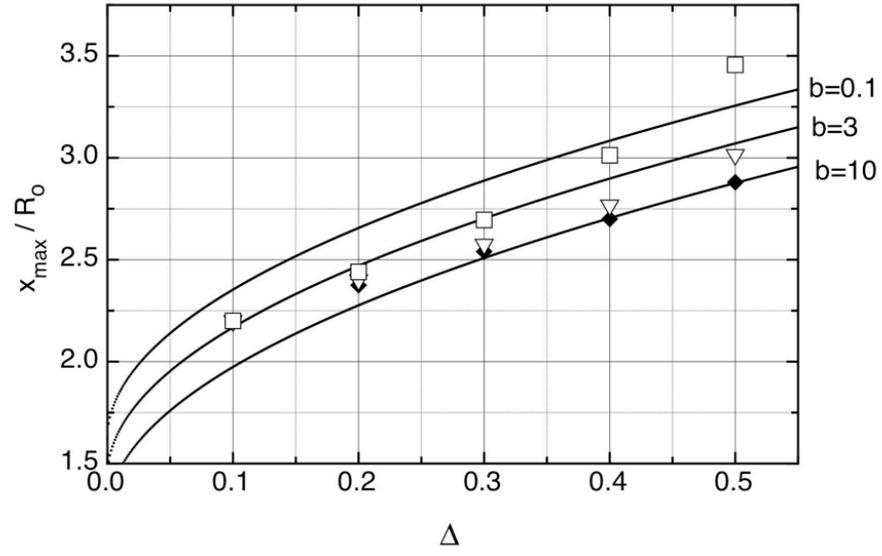

Fig. 6. Comparison of analytical (solid lines) and numerical simulation of maximum particle deviation from the axis: (black dots) $b = 10$, (triangle) $b = 3$, (square) $b = 0.1$.

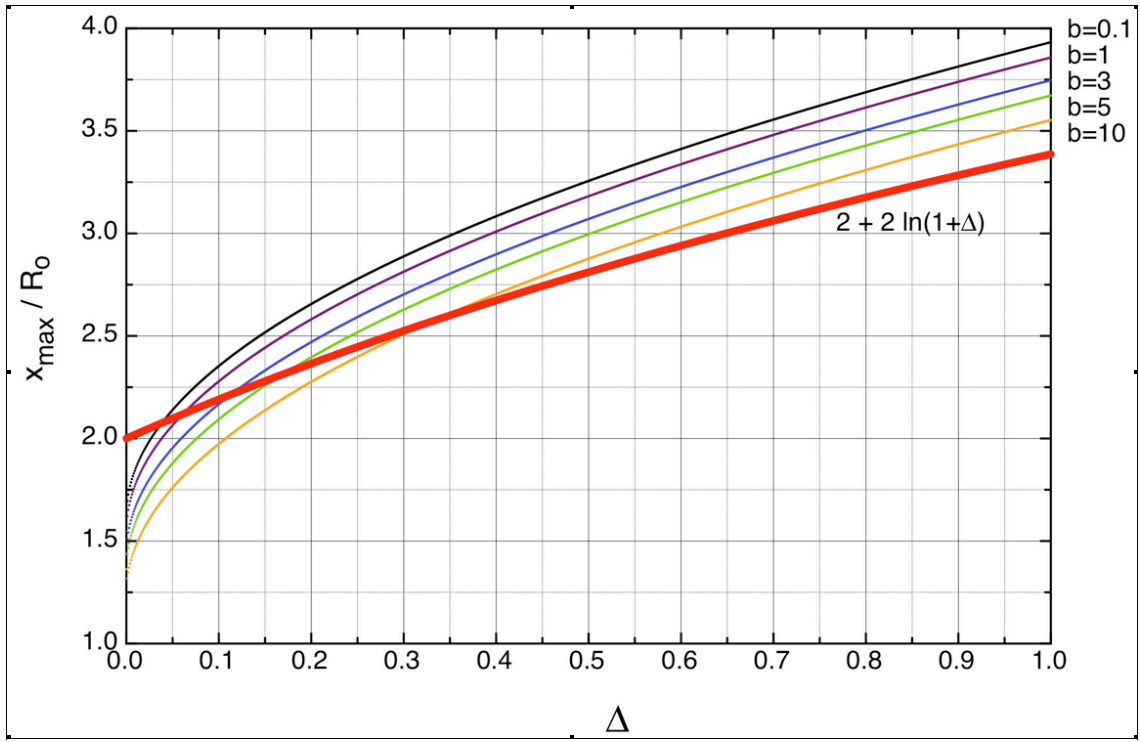

Fig. 7. Maximum deviation of particle from the axis as a function of amplitude of core oscillation.